\documentclass
[preprint,prb,a4paper,amsfonts,amssymb,floats,titlepage,fleqn,footinbib,showpacs,unsortedaddress,twocolumn,10pt,lengthcheck,tightenlines,balancelastpage]{revtex4}%
\usepackage{amsfonts}
\usepackage[fleqn,tbtags]{amsmath}
\usepackage{amssymb}
\usepackage{graphicx}
\usepackage{longtable}%
\providecommand{\U}[1]{\protect\rule{.1in}{.1in}}

\begin{document}
\author{M. ElMassalami}
\affiliation{Instituto de Fisica, Universidade Federal do Rio de Janeiro, 21945-970 Rio de
Janeiro, Brazil}
\date{\today{}}
\title{On the anomalous thermal evolution of the low-temperature, normal-state
specific heat of various nonmagnetic intermetallic compounds.}

\begin{abstract}
The low-temperature normal-state specific heat and resistivity curves of
various nonmagnetic intermetallic compounds manifest an anomalous thermal
evolution. Such an anomaly is exhibited as a break in the slope of the
linearized $C/T$ \textit{versus }$T^{2}$ curve and as a drop in the $R$
\textit{versus} $T$ curve, both at the same $T_{\beta\gamma}$. It is related,
not to a thermodynamic phase transition, but to a Kohn-type anomaly in the
density of states curves of the phonon or electron subsystems. On representing
these two anomalies as additional Dirac-type delta functions, situated
respectively at $k_{B}\theta_{L}$ and $k_{B}\theta_{E}$, an analytical
expression for the total specific heat can be obtained. A least-square fit of
this expression to experimental specific heat curves of various compounds
reproduced satisfactorily all the features of the anomalous thermal evolution.
The obtained fit parameters (in particular the Sommerfeld constant $\gamma
_{0}$ and Debye temperatures $\theta_{D}$) compare favorably with the reported
values. Furthermore, the analysis shows that (i) $T_{\beta\gamma}/\theta
_{L}=0.2\left(  1\pm1/\sqrt{6}\right)  $ and (ii) $\gamma_{0}$ $\propto$
$\theta_{D}^{-2}$; \ both relations are in reasonable agreement with the
experiments. Finally, our analysis (based on the above arguments) justifies
the often-used analysis that treats the above anomaly in terms of either a
thermal variation of $\theta_{D}$ or an additional Einstein mode.

\end{abstract}

\maketitle

\section{Introduction}

The low-temperature, normal-state specific heat of a nonmagnetic intermetallic
is usually described in term of a sum of lattice, $C_{L}$, and electronic,
$C_{E}$, contributions.\cite{Copal-Specificheat} The former is commonly
approximated by the Debye model ($C_{L}=\beta T^{3}$ for $T<<\theta_{D}$ ;
$\theta_{D}$ is the Debye temperature) while the latter by the Sommerfeld
model ($C_{E}=\gamma T$ for $T<<\theta_{F}$; $\theta_{F\text{ }}$ is the Fermi
temperature). Considering that at liquid helium temperatures these two
inequalities are strictly satisfied, then any simultaneous changes in $\beta$
and $\gamma$ would signal a related variation in the spectral features of the
phonon or electron subsystem. Surprisingly, this very $\beta\gamma$-change is
evident as a break in the slope of the normal-state $C/T$ \textit{versus
}$T^{2}$ curve, a break that separates two distinct linearized sections; these
linearly extrapolated upper and lower sections intersect at $T_{\beta\gamma}$
which is taken to be a measure of the energy of this event.

\begin{figure}[th]%
\centering
\includegraphics[
height=6.7348cm,
width=8.1012cm
]%
{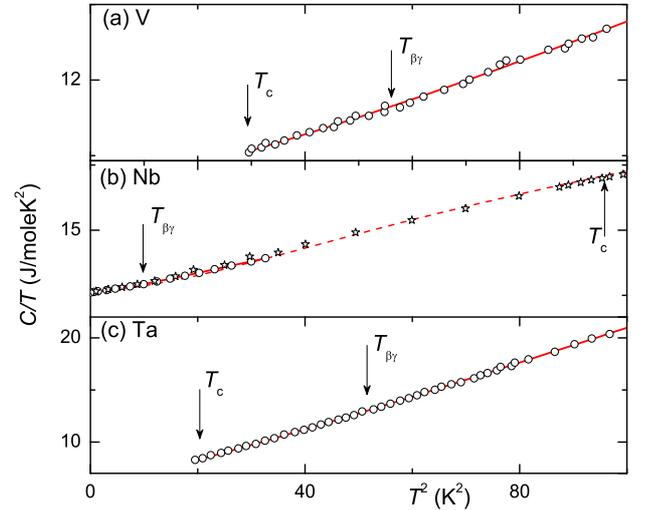}%
\caption{$C/T$ versus $T^{2}$ curves of (a)
V,\cite{Leupold77-Anomaly-SpHeat-Vgroup} (b) Nb \ ($H=1.15$ $T$),
\cite{Dasilva69-Nb-SpHeat,Leupold77-Anomaly-SpHeat-Vgroup} (c) Ta.
\cite{Leupold77-Anomaly-SpHeat-Vgroup}. Lines are fits to Eq. \ref{Eq.C-total}
with parameters as given in Table \ref{Tab.I}. For (b) the solid line is a fit
to the data (circles) of Da Silva \textit{et al.}\cite{Dasilva69-Nb-SpHeat}
while the dashed line to those (stars) of Leupold \textit{et al.}%
\cite{Leupold77-Anomaly-SpHeat-Vgroup}}%
\label{Fig1-spheat-V-Nb-Ta}%
\end{figure}

\begin{figure}[th]%
\centering
\includegraphics[
height=5.3488cm,
width=8.0309cm
]%
{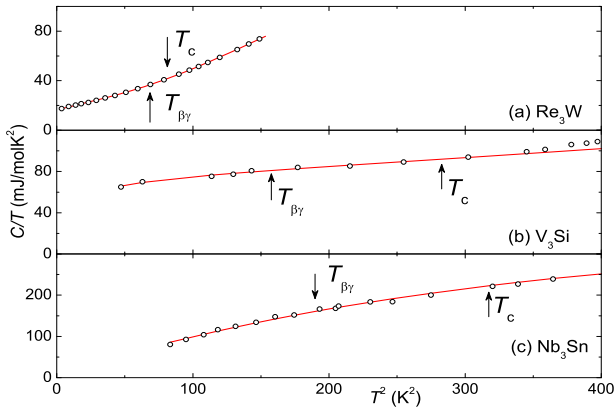}%
\caption{Isofield $C/T$ versus $T^{2}$ curves of the normal-state of (a)
Re$_{3}$W at $H$=7 T, \cite{Yan09-Re3W-SpHeat} (b) V$_{3}$Si at $H$=18 T,
\cite{Stewart84-H-spHeat-A15} (c) Nb$_{3}$Sn at $H$=12.5
T.\cite{Stewart84-H-spHeat-A15} Lines are fits to Eq. \ref{Eq.C-total} with
parameters as given in Table \ref{Tab.I}. The $\beta\gamma$-change of Nb$_{3}%
$Sn and V$_{3}$Si is strongly related to the martensitic transformation, which
reduces the cubic symmetry of the structure into a low-symmetry
one.\cite{Stewart84-H-spHeat-A15} }%
\label{Fig2-SpHeat-Nb3Si-NbSe2}%
\end{figure}

\begin{figure}[tbh]%
\centering
\includegraphics[
height=6.1527cm,
width=8.1012cm
]%
{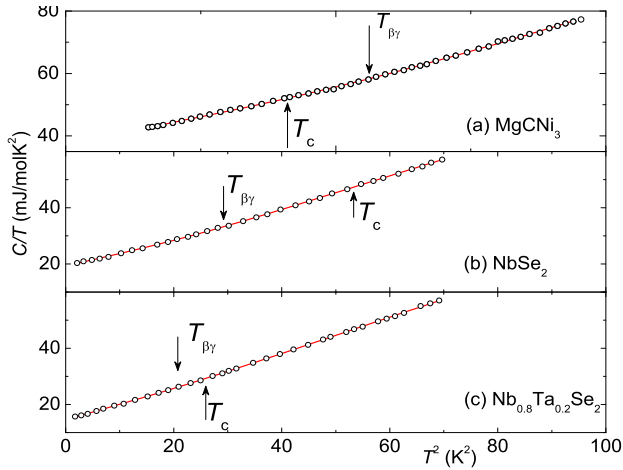}%
\caption{Isofield $C/T$ versus $T^{2}$ curves of (a) MgCNi$_{3}$ at $H$=8 T,
\cite{Lin03-MgCNi3-SpHeat} (b) NbSe$_{2}$ at $H$=6
T,\cite{Nohara99-QP-inVortex-Swave} and (c) Nb$_{0.8}$Ta$_{0.2}$Se$_{2}$ at
$H$=4 T.\cite{Nohara99-QP-inVortex-Swave} Lines are fits to Eq.
\ref{Eq.C-total} with parameters as given in Table \ref{Tab.I}.}%
\label{Fig3-SpHeat-NbSe2-MgCNi3}%
\end{figure}

\begin{figure}[ptb]%
\centering
\includegraphics[
height=5.9067cm,
width=8.1012cm
]%
{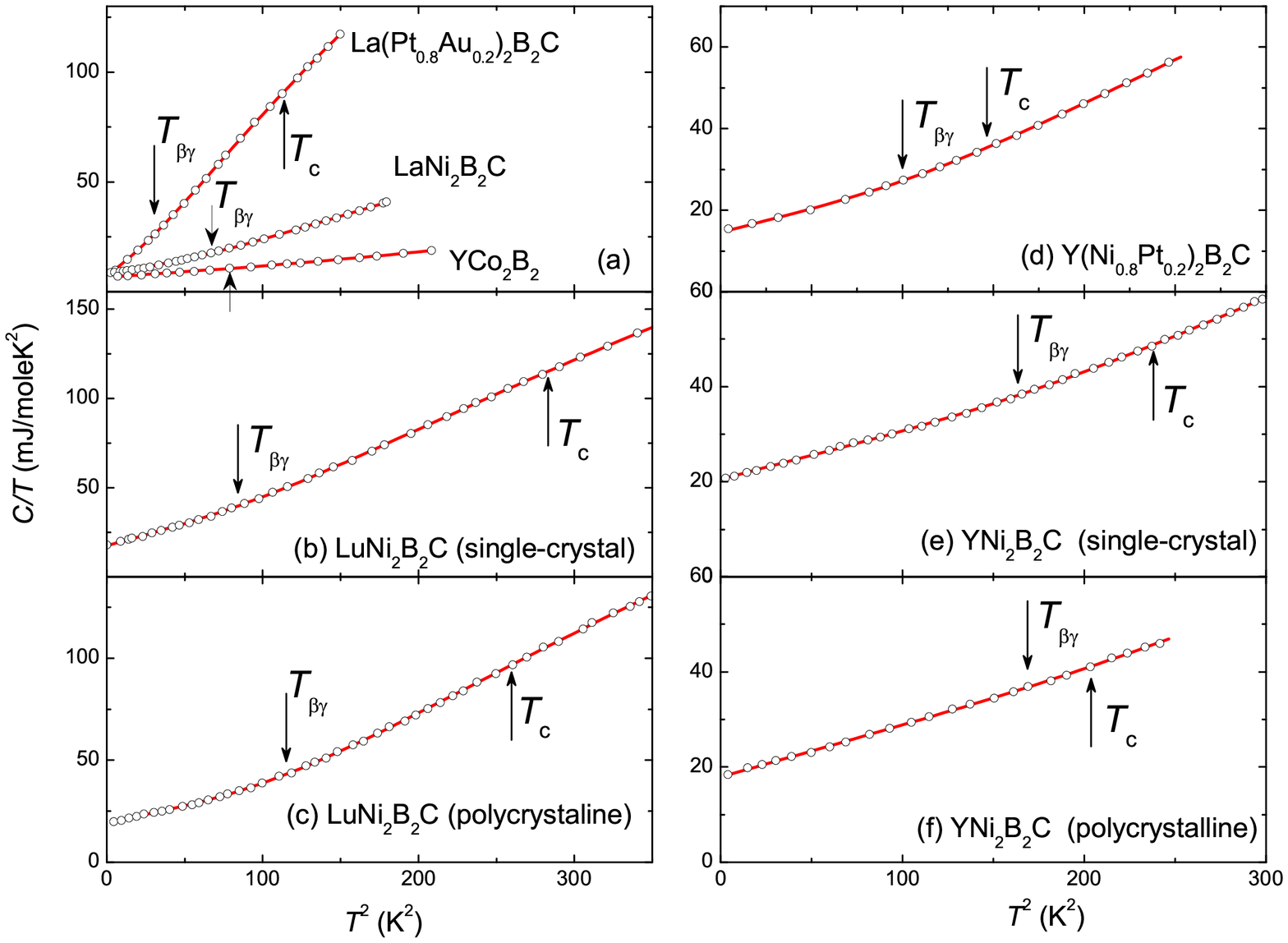}%
\caption{Isofield $C/T$ versus $T^{2}$ curves of (a) La(Pt$_{0.8}$Au$_{0.2}%
$)$_{2}$B$_{2}$C at $H$=9 T while LaNi$_{2}$B$_{2}$C and YCo$_{2}$B$_{2}$ are
at $H$=0 T,\cite{Michor95-C-RNi2B2C} (b) single crystal of LuNi$_{2}$B$_{2}$C
at $H$=8 T, \cite{Nohara97-H-spHeat-LuNi2B2C} (c) polycrystalline LuNi$_{2}%
$B$_{2}$C at $H$=9 T,\cite{Michor95-C-RNi2B2C} (d) Y(Ni$_{0.8}$Pt$_{0.2}%
$)$_{2}$B$_{2}$C at $H_{\Vert c}$=4.5 T, \cite{Nohara99-QP-inVortex-Swave} (e)
single-crystal YNi$_{2}$B$_{2}$C at $H_{\Vert c}$=12 T, and (f)
polycrystalline LuNi$_{2}$B$_{2}$C at $H$=9 T.\cite{Michor95-C-RNi2B2C} Lines
are fits to Eq. \ref{Eq.C-total} with parameters as given in Table
\ref{Tab.I}.}%
\label{Fig4-SpHeat-RNi2B2C}%
\end{figure}

\begin{figure}[tbh]%
\centering
\includegraphics[
height=2.6308in,
width=3.1894in
]%
{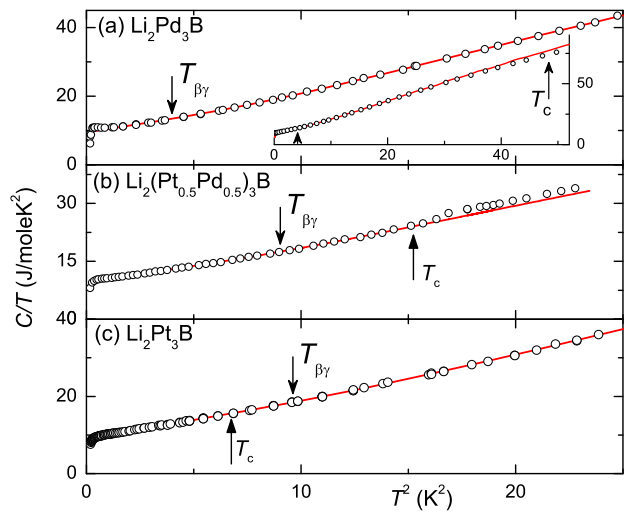}%
\caption{Isofield $C/T$ versus $T^{2}$ curves\ of the normal-state of (a)
Li$_{2}$Pd$_{3}$B at $H=5$ T, (b) Li$_{2}$(Pd$_{0.5}$Pt$_{0.5}$)$_{3}$B at
$H=3$ T, and (c) Li$_{2}$Pt$_{3}$B at $H=3$ T.\cite{07-Li2T3B} Lines are fits
to Eq. \ref{Eq.C-total} with parameters as given in Table \ref{Tab.I}.}%
\label{Fig5-Spheat-Li2PdPt3b}%
\end{figure}

Varieties of materials exhibit this $\beta\gamma$-change: examples include
group V transition metals V, Nb and Ta (Fig. \ref{Fig1-spheat-V-Nb-Ta})
\cite{Morin63-SpHeat-Trans-Metals-Sup,vanderHoeven64-Nb-SpHeat,Dasilva69-Nb-SpHeat,Leupold77-Anomaly-SpHeat-Vgroup}
the Chevrel phases \textrm{PbMo}$_{\mathrm{6}}$\textrm{X}$_{\mathrm{8}}$ (X=S,
Se),\cite{Alekseevskii80-SpHeat-anomaly-Chevrel,Kobayashi82-SpHeat-Mo6Se8} the
$A12$-type Re$_{3}$W,\cite{Yan09-Re3W-SpHeat} the $A15$-type Nb$_{3}$Sn and
V$_{3}$Si (Fig. \ref{Fig2-SpHeat-Nb3Si-NbSe2}%
),\cite{Stewart81-H-spHeat-Nb3Sn,Stewart84-H-spHeat-A15} the layered
NbSe$_{2}$,\cite{Ishikawa71-specificHeat-NbV} the perovskite MgCNi$_{3}$(Fig.
\ref{Fig3-SpHeat-NbSe2-MgCNi3}),\cite{Lin03-MgCNi3-SpHeat} the borocarbides
$R$Ni$_{2}$B$_{2}$C ($R$=Y, La, Lu) (Fig. \ref{Fig4-SpHeat-RNi2B2C}%
),\cite{Michor95-C-RNi2B2C,Nohara97-H-spHeat-LuNi2B2C,Nohara99-QP-inVortex-Swave,03-AFSup-RNi2B2C-Cm}
Li$_{2}$(Pd$_{1-x}$Pt$_{x}$)$_{3}$B ($x$=0, 0.5, 1) (Fig.
\ref{Fig5-Spheat-Li2PdPt3b}),\cite{07-Li2T3B} and Li$_{x}$RhB$_{1.5}$ ($x$=
0.8, 1.0, 1.2).\cite{10-Li-Rh-B-JPSJ} In particular, for Nb$_{3}$Sn, this
$\beta\gamma$-change was shown to be required by the condition of entropy
balance.\cite{Stewart81-H-spHeat-Nb3Sn,Stewart84-H-spHeat-A15}%


Generally, the analysis of the normal-state specific heat is undertaken so as
to obtain $\theta_{D}$\ from $\beta$ and $N(E_{F})$ from $\gamma$ [$N(E_{F})$
is the density of states at the Fermi level, $E_{F}$], then it is not
surprising that such a $\beta\gamma$-anomaly did not attract much attention;
rather it is considered as an undesirable complication that hinders the
precise evaluation of $\theta_{D}$ and $N(E_{F})$. In spite of this
background, some dedicated investigations were carried out even though no
common agreements on the interpretation were
reached.\cite{Morin63-SpHeat-Trans-Metals-Sup,vanderHoeven64-Nb-SpHeat,Dasilva69-Nb-SpHeat,Moore71-Nb-SpHeatanomaly,Leupold77-Anomaly-SpHeat-Vgroup,Stewart81-H-spHeat-Nb3Sn,Pickett82-normal-Nb3Sn,Stewart84-H-spHeat-A15}
Nevertheless, there was a consensus about the following common features.%

\begin{table*}[tbp] \centering
\caption{Normal-state parameters of various compounds as obtained from the least-squares fit of  Eq. \ref{Eq.C-total} to the data of Figs. \ref{Fig1-spheat-V-Nb-Ta}-\ref{Fig5-Spheat-Li2PdPt3b}. $T_{c}$ is the superconducting critical temperature at zero field;  $sc$ ($pc$) denotes a  single-crystal (polycrystalline) sample. Other symbols are explained in the text.}%
\begin{tabular}
[c]{ccccccccc}\hline\hline
Compound & $T_{c}$ & $T_{\beta\gamma}$ & $\theta_{D}$ & $\theta_{L}$ &
$\gamma_{0}$ & b & $T_{\beta\gamma}/\theta_{L}$ & Ref\\
& K & K & K & K & mJ/mol K$^{2}$ &  &  & \\\hline
Nb & 9.28 & 3.11 & 275 & 31.3 & 7.8 & 0.0017 & 0.099 &
\cite{Dasilva69-Nb-SpHeat,Leupold77-Anomaly-SpHeat-Vgroup}\qquad\\
V & 5.4 & 7.5 & 397 & 76.0 & 9.7 & 0.0056 & 0.099 &
\cite{Ishikawa71-specificHeat-NbV,Leupold77-Anomaly-SpHeat-Vgroup}\\
Ta & 4.5 & 7.2 & 238 & 69.0 & 5.4 & 0.0107 & 0.104 & \cite{07-Li2T3B}\\
V$_{3}$Si & 16.8 & 12.6 & 435 & 32.5 & 52.3 & 0.0030 & 0.388 &
\cite{Stewart81-H-spHeat-Nb3Sn,Stewart84-H-spHeat-A15}\\
Nb$_{3}$Sn & 17.8 & 13.7 & 254.3 & 60.1 & 33 & 0.0267 & 0.228 &
\cite{Stewart81-H-spHeat-Nb3Sn,Stewart84-H-spHeat-A15}\\
Re$_{3}$W & 9 & 8.3 & 300.1 & 88.7 & 16.4 & 0.0556 & 0.094 &
\cite{Yan09-Re3W-SpHeat}\\
MgCNi$_{3}$ & 6.4 & 7.5 & 301.0 & 77.5 & 37.4 & 0.0242 & 0.097 &
\cite{Lin03-MgCNi3-SpHeat}\\
NbSe$_{2}$ & 7.3 & 5.4 & 235.5 & 49.2 & 19.3 & 0.0081 & 0.110 &
\cite{Nohara97-H-spHeat-LuNi2B2C,Nohara99-QP-inVortex-Swave}\\
Nb$_{0.8}$Ta$_{0.2}$Se$_{2}$ & 5.1 & 4.5 & 219.6 & 48.0 & 14.6 & 0.0048 &
0.094 & \cite{Nohara97-H-spHeat-LuNi2B2C,Nohara99-QP-inVortex-Swave}\\
YCo$_{2}$B$_{2}$ & 0 & 8.9 & 575.9 & 116.2 & 6.7 & 0.0091 & 0.077 &
\cite{07-Li2T3B}\\
LaNi$_{2}$B$_{3}$C & 0 & 8.2 & 443.4 & 82.1 & 8.0 & 0.0098 & 0.100 &
\cite{07-Li2T3B}\\
La(Pt$_{0.8}$Au$_{0.2}$)$_{2}$B$_{2}$C & 10.7 & 5.6 & 269.7 & 54.4 & 7.1 &
0.0077 & 0.103 & \cite{Michor95-C-RNi2B2C}\\
Y(Pt$_{0.2}$Ni$_{0.8}$)$_{2}$B$_{2}$C & 12.1 & 10 & 461.7 & 106.0 & 14.6 &
0.0258 & 0.094 & \cite{Nohara97-H-spHeat-LuNi2B2C,Nohara99-QP-inVortex-Swave}%
\\
YNi$_{2}$B$_{2}$C (sc) & 15.4 & 12.8 & 481.7 & 136.8 & 20.6 & 0.0410 & 0.094 &
\cite{Nohara97-H-spHeat-LuNi2B2C,Nohara99-QP-inVortex-Swave,03-AFSup-RNi2B2C-Cm}%
\\
YNi$_{2}$B$_{2}$C (pc) & 14.3 & 13 & 463.1 & 164.0 & 17.9 & 0.0709 & 0.079 &
\cite{Michor95-C-RNi2B2C}\\
LuNi$_{2}$B$_{2}$C (pc) & 16.1 & 10.7 & 402.3 & 103.9 & 19.2 & 0.0546 &
0.103 & \cite{Michor95-C-RNi2B2C}\\
LuNi$_{2}$B$_{2}$C (sc) & 16.8 & 9.2 & 358.1 & 95.4 & 17.6 & 0.0279 & 0.096 &
\cite{Nohara97-H-spHeat-LuNi2B2C,Nohara99-QP-inVortex-Swave}\\
Li$_{2}$Pt$_{3}$B & 2.56 & 3.1 & 231.7 & 35.4 & 9.2 & 0.0380 & 0.088 &
\cite{07-Li2T3B}\\
Li$_{2}$Pd$_{1.5}$Pt$_{1.5}$B & 3.9 & 3.0 & 242.4 & 32.2 & 9.8 & 0.0030 &
0.093 & \cite{07-Li2T3B}\\
Li$_{2}$Pd$_{3}$B & 6.95 & 2.0 & 226.2 & 28.9 & 9.4 & 0.0030 & 0.069 &
\cite{07-Li2T3B}\\\hline\hline
\end{tabular}
\label{Tab.I}%
\end{table*}%

First, the manifestation of a $\beta\gamma$-change is evident in various
compounds such as normal intermetallics as well as the normal-state of type II
superconductors: each compound differs strongly from the others in its crystal
structure, chemical composition, electronic properties, and the type of
superconductivity (whether conventional or unconventional). As a result, there
are strong differences in $T_{\beta\gamma}$, in the strength of the event, and
in the curvature at $T_{\beta\gamma}$ (see e.g. Figs.
\ref{Fig2-SpHeat-Nb3Si-NbSe2})].

Second, $\beta$ and $\gamma$ are strongly correlated (the electronic and
phonic degrees of freedom are strongly coupled) and that the trend of this
correlation is not arbitrary: apparently, for the two linearized sections, if
$\gamma_{low}<$ $\gamma_{high}$ then $\beta_{low}$ $>$ $\beta_{high}$ and vice versa.

Third, none of the $C/T$ \textit{versus }$T^{2}$ curves exhibits a
discontinuity or a hysteresis effect at $T_{\beta\gamma}$.

Fourth, in spite of the above-mentioned differences, there are, at least, two
ingredients common to most of the studied materials: a relatively strong
electron-phonon coupling and a lifting of spin-degeneracy at $E_{F}$, say, by
an applied magnetic field (as in, e.g., conventional superconductors) or by an
anisotropic spin orbit coupling ASOC interaction (as in, e.g.,
non-centro-symmetric Li$_{2}$(Pd$_{1-x}$Pt$_{x}$)$_{3}$B\ and Li$_{x}%
$RhB$_{1.5}$ superconductors). Curiously, although $H$ ($>H_{c2}$) is
necessary for the quench of the superconductivity and for the lift of spin
degeneracy, it has no influence on $T_{\beta\gamma}$, $\beta$ or $\gamma
$%
.\cite{Stewart81-H-spHeat-Nb3Sn,Stewart84-H-spHeat-A15,Michor95-C-RNi2B2C,Nohara97-H-spHeat-LuNi2B2C,Nohara99-QP-inVortex-Swave}
Fifth, $T_{\beta\gamma}$ can be controlled by substitution.

Various investigators attributed this $\beta\gamma$-change to an anomaly in
the
phonon\cite{Moore71-Nb-SpHeatanomaly,Leupold77-Anomaly-SpHeat-Vgroup,Pickett82-normal-Nb3Sn}
or electron\cite{Leupold77-Anomaly-SpHeat-Vgroup} density of states
DOS\ curves.\cite{Leupold77-Anomaly-SpHeat-Vgroup}\ Attempts were made to
supplement the Debye model by assuming either an additional Einstein
mode\cite{Pickett82-normal-Nb3Sn} or a variation of the effective $\theta_{D}%
$:\cite{Copal-Specificheat} for the latter case, it is often argued that,
based on a typical phonon spectrum, $\theta_{D}(T)$ must decrease as the
thermal energy is raised towards the first peak of the phonon
spectrum.\cite{Kittel-ISSP}%

Two investigations deserve a special mention: First, Stewards \textit{et
al}.\cite{Stewart81-H-spHeat-Nb3Sn,Stewart84-H-spHeat-A15} attributed the
$\beta\gamma$-change in Nb$_{3}$Sn and V$_{3}$Si to the inability of some
acoustic phonons to decay through the creation of electron-quasiparticle
pairs: the opening of the superconducting gap $\Delta_{s}$ leads to an abrupt
change in the lifetimes of phonons with energies less than $2\Delta_{s}$%
($T)$.\cite{Axe73-ND-Nb3Sn,Stewart81-H-spHeat-Nb3Sn} The net results is that
$C_{L}+C_{e}$ would be modified. Second, Moore and
Paul\cite{Moore71-Nb-SpHeatanomaly} as well as Leupold\textit{ et
al.}\cite{Leupold77-Anomaly-SpHeat-Vgroup} assumed that the $\beta\gamma
$-deviation in group V transition metals is related to a Kohn-type anomaly in
the phonon (or electron) spectra. They demonstrated that such a deviation can
be satisfactorily reproduced if one represents the DOS of such an anomaly by a
Dirac delta-function and calculates analytically the total specific heat.%

Although these two approaches addressed adequately some features of the
$\beta\gamma$-deviation, however serious questions are not fully addressed:
e.g. (i) the origin of both $T_{\beta\gamma}$ and the strong and systematic
correlation between $\beta$ and $\gamma$, (ii) the identification of the role
played by both the electron-phonon couplings and the dielectric properties of
the investigated intermetallics, (iii) the observations that the $\beta\gamma
$-change occurs in the absence of superconductivity (as in normal
intermetallics) and that some superconductors, even with an established
$\Delta_{s}$($T)$, do not show this $\beta\gamma$-change.

In this work, we present a systematic analysis of the above-mentioned
anomalous thermal evolution of the specific heat (and resistivity) of various
compounds. The dielectric properties of the compounds under study will be
taken into consideration.\ The above-mentioned approaches of Stewards
\textit{et al}.\cite{Stewart81-H-spHeat-Nb3Sn,Stewart84-H-spHeat-A15} and
Moore and Paul\cite{Moore71-Nb-SpHeatanomaly} will be partially modified,
extended, and generalized. The obtained analytical expression for the total
specific heat (electron plus phonon) is shown to reproduce satisfactorily the
studied experimental curves.%

\section{Theoretical Background}

\subsection{Lattice specific heat}

An anomaly, such as a Kohn type, in the phonon DOS can be represented, as done
in earlier
investigations,\cite{Moore71-Nb-SpHeatanomaly,Leupold77-Anomaly-SpHeat-Vgroup}
by a Dirac delta function. Then%
\begin{equation}
D(E)=a.3N_{L}.E^{2}+b.3N_{L}.\delta(E-E_{L}) \label{Eq.DOS-Lat}%
\end{equation}
where $N_{L}=rN_{A}$ is the number of total atoms, $N_{A}$ is the Avogadro
number, and $E_{L}=$ $k_{B}\theta_{L}$ is the energy at which the Dirac-type
anomaly is situated. $a\ $and $b$ represent the fractional weights and are
related by the normalization condition $%
{\displaystyle\int\limits_{0}^{E_{D}}}
D(E).dE=3rN_{A}$ which gives $a=3(1-b)/\left(  k\theta_{D}\right)  ^{3}$. The
lattice specific heat is then \begin{widetext}%
\begin{align}
&  C_{L}(T)=%
{\displaystyle\int\limits_{0}^{E_{D}}}
E.D(E)\frac{1}{\exp(\frac{E}{k_{B}T})-1}dE=%
{\displaystyle\sum\limits_{n=1}^{\infty}}
{\displaystyle\int\limits_{0}^{E_{D}}}
E.D(E).\exp(-\frac{nE}{k_{B}T})dE=\frac{12}{5}\pi^{4}N_{L}.k_{B}\left(
\frac{T}{\theta_{D}}\right)  ^{3}\xi=\beta_{0}.T^{3}\xi\label{Eq.C-lattice}\\
&  \xi=(1-b)\left\{  1-\frac{15}{4\pi^{4}}\left(  \frac{\theta_{D}}{T}\right)
^{4}L_{0}(Z_{D})-\frac{15}{\pi^{4}}\left(  \frac{\theta_{D}}{T}\right)
^{3}L_{1}(Z_{D})-\frac{45}{\pi^{4}}\left(  \frac{\theta_{D}}{T}\right)
^{2}L_{2}(Z_{D})-\frac{90}{\pi^{4}}\left(  \frac{\theta_{D}}{T}\right)
L_{3}(Z_{D})-\frac{90}{\pi^{4}}L_{4}(Z_{D})\right\}  \nonumber\\
&  +b\left\{  \frac{5}{4\pi^{4}}\left(  \frac{\theta_{D}}{T}\right)
^{3}\left(  \frac{\theta_{L}}{T}\right)  ^{2}L_{-1}(Z_{L})\right\}
\label{Eq.Lattice-correction}%
\end{align}
\end{widetext}where $L_{i}(Z)$ is the polylogarithm function; $Z_{D}%
=\exp(-\theta_{D}/T)$; $Z_{L}=\exp(-\theta_{L}/T)$. The expression with $b=0$
is the usual Debye contribution (for numerical calculation, this form is much
better than the one given in solid-state text
books).\cite{Copal-Specificheat,Kittel-ISSP} The Debye $T^{3}$ expression is
obtained when $T\rightarrow0$ [$\xi$ acts as a $T$-dependent correction
factor, see Fig. \ref{Fig6-Simulation}(a)] while the Dulong-Petit value is
reached when $T>\theta_{D}$. Eq. \ref{Eq.C-lattice} justifies the often-used
procedure of treating the $\beta\gamma$-deviation as a thermal variation of an
effective temperature-dependent $\theta_{D}^{\prime}=\theta_{D}/\xi^{1/3}$:
indeed Fig. \ref{Fig6-Simulation} (b) reproduces the often-observed thermal
variation of $\theta_{D}^{\prime}$.

The low-temperature limit of Eq. \ref{Eq.C-lattice} reproduces the expressions
of Moore and Paul\cite{Moore71-Nb-SpHeatanomaly} as well as that of
Leupold\textit{ et al.}\cite{Leupold77-Anomaly-SpHeat-Vgroup} Evidently, the
introduction of that anomaly leads to both a Debye-type contribution with
weight factor ($1-b$) and an Einstein-type contribution with frequency
$\omega_{L}$ and weight factor $b$. The latter conclusion justifies the
often-used practice of adding an Einstein contribution to the specific heat of
such intermetallic systems.\cite{Pickett82-normal-Nb3Sn}

\subsection{Electronic specific heat}

Along similar lines, a DOS curve of an electron subsystem with an additional
Dirac delta function can be represented
as\cite{Leupold77-Anomaly-SpHeat-Vgroup}%
\begin{equation}
N(E)=c.N_{E}.E^{\frac{1}{2}}-d.N_{E}.\delta(E-E_{E}) \label{Eq.DOS-el}%
\end{equation}
where $N_{E}=xN_{A}$ is the total number of conduction electrons and $E_{E}=$
$k_{B}\theta_{E}$ denotes the position of the Dirac-type anomaly. $c$\ and $d$
are fractional weights that are related by the normalization condition%
\begin{equation}%
{\displaystyle\int\limits_{0}^{\infty}}
dE.N(E)/\left(  \exp(\frac{E-E_{F}}{k_{B}T})+1\right)  =xN_{A}.
\label{Eq.-Electron-Number}%
\end{equation}
This gives
\[
c\cong\frac{1+d/\left[  1+\exp((\theta_{E}-\theta_{F})/T)\right]  }{\frac
{2}{3}E_{F}^{3/2}\left[  1+(\pi T/\sqrt{8}\theta_{F})^{2}\right]  }.
\]
As the electron number is independent of temperature, then the thermal rate of
the chemical potential is $-\pi^{2}k_{B}^{2}.T/6E_{F}$ as $T\rightarrow0$.

Using Sommerfeld expansion in the expression of the electronic energy and
taking the derivative with respect to temperature, one obtains
\begin{widetext}
\begin{align}
C_{E}(T)  &  \cong\frac{\pi^{2}N_{E}k_{B}T}{2\theta_{F}}\eta=\gamma
_{0}T\eta\label{Eq.C-electron}\\
\eta &  =1+\frac{d}{1+\exp((\theta_{E}-\theta_{F})/T)}-d\frac{1}{2\pi^{2}%
T^{3}}\theta_{E}\theta_{F}\left(  \theta_{E}-\theta_{F}\right)
\text{sech}^{2}((\theta_{E}-\theta_{F})/2T) \label{Eq.electron-correction}%
\end{align}
\end{widetext}$\eta$ is a $T$-dependent correction factor, see Fig.
\ref{Fig6-Simulation}(c). Eq.\ref{Eq.C-electron} is a sum of three
contributions: the first is the usual Sommerfeld expression ($d=0$) while the
second and third terms are related to the Dirac delta function.

\subsection{Total specific heat (Lattice plus electron)}

\begin{figure}[tbh]%
\centering
\includegraphics[
height=6.1cm,
width=8.1012cm
]%
{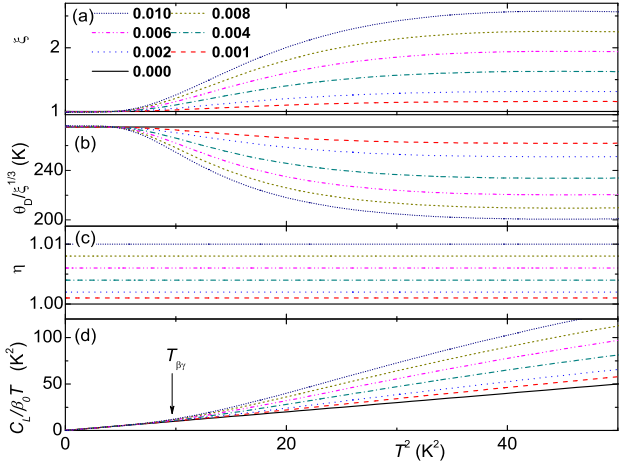}%
\caption{(Color online) The evolution of (a) $\xi$, (b) the effective
$\theta_{D}^{\prime}=\theta_{D}/\xi^{1/3}$, (c) $\eta$, and (d) $\xi
.T^{2}=C_{L}/\beta_{0}T$ versus $T^{2}$ for various values of $b$ ($b=d$). The
following values were used (typical for
Nb):\cite{Leupold77-Anomaly-SpHeat-Vgroup} $\theta_{D}=275$ K, $\theta
_{L}=\theta_{E}=$33 K, and $\gamma_{0}=7.8$ mJ/moleK$^{2}$ (see Table
\ref{Tab.I}). For $T<\theta_{D}$, the thermal variation of $\eta$ is almost
constant while $\xi$ varies sharply above $T_{\beta\gamma}$. Evidently the
slope break at $T_{\beta\gamma}$ is due to the thermal variation of $\xi
.T^{2}$ = $C_{L}/\beta_{0}T$ = $\left[  C_{tot}-\gamma_{0}(1+d)T\right]
/\beta_{0}T$\ ($\beta_{0}$ and $\gamma_{0}$ are $T-$independent).}%
\label{Fig6-Simulation}%
\end{figure}

From Eqs. \ref{Eq.C-lattice} and \ref{Eq.C-electron}, the total $C/T$
\textit{versus }$T^{2}$ can be presented in the familiar form
\begin{equation}
C/T=\gamma_{0}.\eta+\beta_{0}.\xi.T^{2}=\gamma+\beta T^{2} \label{Eq.C-total}%
\end{equation}
For $T<\theta_{D}<<\theta_{F}$, $\theta_{E}$ $<<\theta_{F}$, and $\theta
_{L}<\theta_{D}$, both linear Sommerfeld and cubic Debye approximation are
obtained but with effective $\gamma$ and $\beta$. Fig. \ref{Fig6-Simulation}%
(d) shows that, within this temperature range, the thermal evolution of $\eta$
and $\xi$ (consequently that of $\gamma$ and $\beta$) leads to the
experimentally observed two limiting behaviors. Analytically, the
low-temperature ($\theta_{E},\theta_{L}>T_{\beta\gamma}>T\rightarrow0$) limit
is%
\begin{align}
\gamma_{low}  &  =\gamma_{0}(1+d)\label{Eq.gamaLow}\\
\beta_{low}  &  =\beta_{0}(1-b) \label{Eq.betaLow}%
\end{align}
while the high-temperature ($\theta_{E},\theta_{L}>T>T_{\beta\gamma}$) limit
is%
\begin{align}
\gamma_{high}  &  \approx\gamma_{0}\left[  1+d\left(  1+\frac{4}{\pi^{2}}%
\frac{\theta_{E}}{\theta_{F}}X^{3}\text{sech}^{2}(-X)\right)  \right]
\label{Eq.gamaHigh}\\
\frac{\gamma_{high}}{\gamma_{low}}  &  =1+\frac{4d\theta_{E}X^{3}%
\text{sech}^{2}(-X)}{\pi^{2}(1+d)\theta_{F}}\label{Eq.gamaRatio}\\
\beta_{high}  &  \approx\beta_{0}\left[  1-b\left(  1-\frac{5}{4\pi^{4}%
}\left(  \frac{\theta_{D}}{\theta_{L}}\right)  ^{3}\right)  \right]
\label{Eq.betaHigh}\\
\frac{\beta_{high}}{\beta_{low}}  &  =1+\frac{5b}{4\pi^{4}(1-b)}\left(
\frac{\theta_{D}}{\theta_{L}}\right)  ^{3} \label{Eq.betaRatio}%
\end{align}
where $X=\theta_{F}/2T$. The thermal variation of $\gamma$\ (Eqs.
\ref{Eq.electron-correction}, \ref{Eq.gamaLow}, \ref{Eq.gamaHigh}) as compared
to that of $\beta$ (Eqs. \ref{Eq.Lattice-correction}, \ref{Eq.betaLow},
\ref{Eq.betaHigh}) is extremely small: as a consequence the thermal variation
of Eq. \ref{Eq.C-total} is predominantly governed by that of $\beta$. This
explains the reported success in analyzing the specific heats of group V
transition metals in terms of\ $C/T=\gamma_{low}+\beta_{0}\xi T^{2}%
$.\cite{Moore71-Nb-SpHeatanomaly,Leupold77-Anomaly-SpHeat-Vgroup} Then the
intensity and extent of slope break are related mainly to $\xi$ (Eq.
\ref{Eq.Lattice-correction}) (extremely weak dependence on $\theta_{E}$,
$N_{E}$, and $d$ is expected). This conclusion allows us to define
$T_{\beta\gamma}$ as the point of inflection in the $\xi(T)$ curve [see Eq.
\ref{Eq.Lattice-correction} and Fig. \ref{Fig6-Simulation}(a)]. Then the
solution of $\partial^{2}\xi/\partial T^{2}$=0 gives
\begin{equation}
T_{\beta\gamma}=\frac{\theta_{L}}{5}\left(  1\pm\frac{1}{\sqrt{6}}\right)
\cong0.12\theta_{L},\text{ }0.28\theta_{L} \label{Eq.Tbg-ThetaRatio}%
\end{equation}
These two calculated ratios\ of $\frac{T_{\beta\gamma}}{\theta_{L}}$ are in
reasonable agreement with the experimentally determined values of Nb which
manifests two slope
breaks:\cite{Chou58-SpHeat-Nb,Morin63-SpHeat-Trans-Metals-Sup,vanderHoeven64-Nb-SpHeat,Dasilva69-Nb-SpHeat,Moore71-Nb-SpHeatanomaly,Leupold77-Anomaly-SpHeat-Vgroup}
one at $T_{\beta\gamma}^{\prime}$ = 3.11 K while the other at $T_{\beta\gamma
}^{^{\prime\prime}}$ =10.3 K. Using $\theta_{L}=31.3$ K (see Table
\ref{Tab.I}), then the calculated $T_{\beta\gamma}^{\prime}$ = 3.8 K while
$T_{\beta\gamma}^{^{\prime\prime}}$ =8.8 K (the experimental $\frac
{T_{\beta\gamma}}{\theta_{L}}$ ratios are, respectively, 0.10 and 0.33). It is
noted that the high $T_{\beta\gamma}$-event (showing the highest discrepancy)
was not reported\cite{Leupold77-Anomaly-SpHeat-Vgroup} for the isomorphous Ta
and V suggesting that it might be unique to Nb.

Equations \ref{Eq.betaHigh}-\ref{Eq.betaRatio} predict a positive curvature
and $\beta_{high}>\beta_{low}$ whenever\ $b>0$: this is consistent with the
features of all analyzed curves in Figs. \ref{Fig1-spheat-V-Nb-Ta}%
-\ref{Fig5-Spheat-Li2PdPt3b} except those of V$_{3}$Si\ and Nb$_{3}$Sn; these
indicate that $b<0$.

The above arguments, in particular Fig. \ref{Fig6-Simulation}, emphasize that,
the slope break (whether smooth or sharp) is not a phase transition, rather it
is a consequence of the thermal evolution of $\xi$ (and,\ to a lesser extent,
$\eta$): indeed no thermodynamic phase transition in borocarbides was reported
in the extensively measured low-$T$
magnetoresistivity,\cite{Miyamoto97-YNi2B2C-MagRes,Chu00-YNi2B2C-MagRes,Fisher97-RNi2B2C-magnetoresistivity,Ranthnayaka97-(YLu)Ni2b2C}%
, thermopower\cite{Ranthnayaka97-(YLu)Ni2b2C}, thermal
conductivity,\cite{Cao00-thermal-conductivity-RNi2B2C}
Hall,\cite{Narozhnyi99-(YLu)Ni2B2C-Hall} and thermal
expansion\cite{Budko06-YNi2B2C-Magnetostriction} properties.

\subsection{The correlation of $\theta_{D}$ ($\beta$) and $\gamma$}

A correlation between $\beta$ and $\gamma$ can be obtained if we consider the
low-temperature specific heat of these intermetallics as being due to
Sommerfeld-type electrons and longitudinal acoustic Debye-type phonons. As the
total dielectric function (electrons plus ions) of these longitudinal modes
must be zero, then the sound velocity, as $\left\vert \vec{k}\right\vert
\rightarrow0$, would be $\upsilon_{s}=\upsilon_{F}\sqrt{m/3M}$, where
$\upsilon_{F}$, $m$, and $M$ denote, respectively, the Fermi velocity, the
electronic mass, and ionic mass.\cite{Kittel-ISSP} Inserting this
$\upsilon_{s}$ into $\theta_{D}$ (as obtained from the Debye model) and
replacing $\upsilon_{F}$ by $\gamma$ (as obtained from Sommerfeld model) one
gets the correlation of $\gamma$ and $\theta_{D}$:%
\begin{equation}
\gamma=\left(  \frac{3\pi^{4}h^{6}N_{A}^{5}}{2}\right)  ^{1/3}\frac{1+d}%
{M}\left(  \frac{x^{3}r^{2}}{V_{m}^{2}}\right)  ^{1/3}\frac{1}{\theta_{D}^{2}%
}\label{Eq.Gamma-Theta}%
\end{equation}
where $V_{m}$ is the molar volume. For $N_{E}=N_{L}$ (or $x=r$) and
$\theta_{D}$ as obtained from Table \ref{Tab.I}, Eq. \ref{Eq.Gamma-Theta}
gives $\gamma_{cal}$ having the same order of magnitude as the experimentally
determined $\gamma$.

\subsection{The $T_{\beta\gamma}$-event as a resistivity drop}

The $T_{\beta\gamma}$-events in Nb,\cite{Leupold77-Anomaly-SpHeat-Vgroup} Fig.
\ref{Fig1-spheat-V-Nb-Ta}, and its Nb$_{1-x}Y_{x}$ ($Y=$ Ti, W) alloys were
confirmed by the manifestation of a resistivity drop, at the same
$T_{\beta\gamma}$.\cite{Chopra72-Resistivity-Nb} Based on the above-mentioned
arguments, this drop can be interpreted along the following two lines: (i) The
electronic concentration $n$ in the neighborhood of $T_{\beta\gamma}$\ varies
as $n_{0}/\eta^{3/2}$ ($n_{0}$ is the electronic concentration at $d=0$). Then
an increase in $n$ below $T_{\beta\gamma}$ would lead to such a resistivity
drop. (ii) The temperature-dependent resistivity is usually approximated by
the Bloch--Gr\"{u}neisen
expression:\cite{Allen96-Quantum-Theory-of-Real-Materials}%
\begin{equation}
\rho(T)-\rho_{0}=(4\pi)^{2}(\lambda\omega_{p}^{-2})\omega_{D}\left(  \frac
{2T}{\theta_{D}}\right)  ^{5}%
{\displaystyle\int\nolimits_{0}^{\frac{\theta_{D}}{2T}}}
\frac{x^{5}}{\sinh(x)^{2}}dx \label{Eq.ResBlochGruneisen}%
\end{equation}
where $\rho_{0}$ is the temperature-independent contribution, $\lambda$ is the
electron-phonon coupling, and $\omega_{p}$ is the Drude plasma frequency
($\omega_{p}^{2}=4\pi e^{2}n/m$; factors have their usual meaning). Then an
increase in $\theta_{D}$ (a drop in $\beta$) below $T_{\beta\gamma}$ would be
manifested as a decrease in the resistivity.%

\section{Analysis and discussion}

All the curves\ reported in Figs. \ref{Fig1-spheat-V-Nb-Ta}%
-\ref{Fig5-Spheat-Li2PdPt3b} were least-squares analyzed with Eq.
\ref{Eq.C-total} assuming $\theta_{E}=\theta_{L}$ and $d=b$: these reasonable
simplifications would not influence any of the conclusions drawn from this
analysis since, as mentioned above, the dominant thermal variation is due to
that of $\xi$.

As can be observed in Figs. \ref{Fig1-spheat-V-Nb-Ta}%
-\ref{Fig5-Spheat-Li2PdPt3b}, there is a satisfactory agreement between theory
(solid lines) and experiment (symbol). The fit parameters are shown in Table
\ref{Tab.I}: comparison with the published values suggests that the obtained
$\gamma_{0}$ is as expected but $\theta_{D}$ is slightly different: this is
attributed to the difference in the criterion for its determination. In
addition, Table \ref{Tab.I} shows that (i) $b$ is extremely small as expected,
(ii) $\gamma_{0}$ is correlated to $\theta_{D}$ through Eq.
\ref{Eq.Gamma-Theta}, and (iii) the $T_{\beta\gamma}/\theta_{L}$ ratio is in
accord with Eq. \ref{Eq.Tbg-ThetaRatio}: in fact this ratio can be used to
classify the studied compounds: those with $T_{\beta\gamma}/\theta
_{L}\thickapprox\left(  1+1/\sqrt{6}\right)  /5$ (such as Nb$_{3}$Sn, V$_{3}%
$Si) have $b<0$ and a negative curvature while all the other with
$T_{\beta\gamma}/\theta_{L}\thickapprox\left(  1-1/\sqrt{6}\right)  /5$ have
$b>0$ and a positive curvature.

Figure \ref{Fig1-spheat-V-Nb-Ta} shows that $T_{\beta\gamma}<T_{c}$ for Nb,
while $T_{\beta\gamma}>T_{c}$\ for Ta and
V:\cite{Leupold77-Anomaly-SpHeat-Vgroup} thus $T_{\beta\gamma}$ (or
$\theta_{L}$) is not strictly correlated with the pairing potential.
Furthermore $T_{\beta\gamma}$ does not show any systematic correlation with
the atomic weight: indeed a Kohn-type anomaly is not necessarily related to
the atomic masses. Such a nonsystematic feature is evident also in the
normal-state curves of the $A15$
members\cite{Stewart84-H-spHeat-A15,Guritanu04-Nb3Sn-SpHeat} Nb$_{3}$Sn,
V$_{3}$Si, and Re$_{3}$W (Fig. \ref{Fig2-SpHeat-Nb3Si-NbSe2}).

Figure \ref{Fig3-SpHeat-NbSe2-MgCNi3} shows a $\beta\gamma$-change in
MgCNi$_{3}$ (with $T_{\beta\gamma}>T_{c}$)\cite{Lin03-MgCNi3-SpHeat} as well
as in NbSe$_{2}$ and Nb$_{0.8}$Ta$_{0.2}$Se$_{2}$ ( $T_{\beta\gamma}<T_{c}%
$).\cite{Nohara99-QP-inVortex-Swave} It is noted that a 20\% Ta substitution
leads to a decrease in $T_{\beta\gamma}$. On the other hand, Fig.
\ref{Fig4-SpHeat-RNi2B2C} indicates that $b$ is reasonably strong for the
$R$Ni$_{2}$B$_{2}$C compounds indicative of a stronger anomalous contribution.
Interestingly, the $\beta\gamma$-event is evident also in the
nonsuperconductors LaNi$_{2}$B$_{2}$C and YCo$_{2}$B$_{2}$: $b$ (also\ $d$)
does not depend on the pairing potential. In addition, $T_{\beta\gamma}$ of
$R$Ni$_{2}$B$_{2}$C varies over a wider range, from 5 to 13 K even though
$N$($E_{F}$) does not:\cite{Muller01-interplay-review} $\theta_{L}$ does not
depend on $N$($E_{F}$). For all $R$Ni$_{2}$B$_{2}$C, $T_{\beta\gamma}$
($<T_{c}$ when applicable) does manifest a dependence (though unsystematic) on
the sample format (single- or poly-crystals) and on the type of isovalent ions
$R^{3+}$=Y$^{3+}$, La$^{3+}$, Lu$^{3+}$.

Most of the above-mentioned superconductors are type-II with strong or
intermediate-to-strong couplings. Nonetheless, a similar $\beta\gamma$-change
is evident in non-centro-symmetric superconductors such as Re$_{3}$W (Fig
\ref{Fig2-SpHeat-Nb3Si-NbSe2}(a)),\cite{Yan09-Re3W-SpHeat,Zuev07-Re3W}
Li$_{2}$(Pd$_{1-x}$Pt$_{x}$)$_{3}$B ($x$=0, 0.5, 1) (Fig.
\ref{Fig5-Spheat-Li2PdPt3b}),
\cite{07-Li2T3B,Yuan-Swave-ST-SS-Li2PtPD3B,Nishiyama07-ST-broken-inversion-symm-Li2Pt3B}
and Li$_{x}$Rh$_{1.5}$ ($x$= 0.8, 1.0, 1.2).\cite{10-Li-Rh-B-JPSJ} The space
groups of these unconventional superconductors have no inversion symmetry
operator and as such there is a relatively strong ASOC\ interaction which has
the effect of lifting the spin-degeneracy at the Fermi
surface.\cite{Sigrist91-Uncon-SUC,Bauer05-CePt3Si-SUp-Norm,Sigrist07-noncon-Sup-nonCentroSym,Frigeri04-Inverson-symmetry-SUC}%
\ Then, for, say, Li$_{2}$(Pd$_{1-x}$Pt$_{x}$)$_{3}$B, an increase in $x$ is
accompanied by an increase in the ASOC\ interaction, in the effectiveness of
degeneracy lifting, and in $T_{\beta\gamma}$: for lower ASOC, $T_{\beta\gamma
}<T_{c}$ while for higher\ ASOC, $T_{\beta\gamma}>T_{c}$.

Equations \ref{Eq.gamaRatio} and \ref{Eq.betaRatio}, each being related to an
independent subsystem, suggest an absence of appreciable correlation between
$\left(  \frac{\beta_{high}}{\beta_{low}}\right)  _{cal}$ and $\left(
\frac{\gamma_{high}}{\gamma_{low}}\right)  _{cal}$: in fact the former depends
on $b$ and $\frac{\theta_{D}}{\theta_{L}}$ while the latter on $d$,
$\frac{\theta_{E}}{\theta_{F}}$, and $\frac{\theta_{F}}{T}$; only for the
$T\rightarrow0$ limit, $\left(  \frac{\gamma_{high}}{\gamma_{low}}\right)
_{cal}$ is $T$-independent but then there is no correlation since it tends to
1. On the other hand, there is a strong correlation between the experimentally
determined $\left(  \frac{\beta_{high}}{\beta_{low}}\right)  _{\exp}$ and
$\left(  \frac{\gamma_{high}}{\gamma_{low}}\right)  _{\exp}$ (obtained from
the two linearized sections). This apparent inconsistency can be clarified if
we note that (i) $\beta_{low}$\ (Eq. \ref{Eq.betaLow})\ and $\gamma_{low}%
\ $(Eq. \ref{Eq.gamaLow}) are equal to the experimentally determined ones and
both are correlated by Eq. \ref{Eq.Gamma-Theta}, (ii) $\beta_{high}$ (Eq.
\ref{Eq.betaHigh}) is the same as the experimentally determined value but
$\gamma_{high}$ (Eq. \ref{Eq.gamaHigh}) is not the same as the experimentally
determined value due to the involvement of $\xi(T)$: accordingly Eq.
\ref{Eq.Gamma-Theta} can not be used to related the experimentally determined
$\theta_{D,high}$ and $\gamma_{high}$.

It seems that for the cases where $T_{\beta\gamma}<T_{c}$, the zero-field
superconducting state masks the $T_{\beta\gamma}$-event, however, the latter
can be recovered if the superconductivity is quenched with, say, $H>H_{c2}$.
In general, the $T_{\beta\gamma}$-event, if dominated by phonon contributions,
is $H-$independent but can be influenced by substitution. Two type of
substitution-induced influences can be identified: in the first, an increase
in doping leads to a decrease in $T_{\beta\gamma}$ such as in the case of
Y(Ni$_{1-x}$Pt$_{x}$)$_{2}$B$_{2}$C, Nb$_{1-x}$V$_{x}$%
,\cite{Ishikawa71-specificHeat-NbV} and Nb$_{0.8}$Ta$_{0.2}$Se$_{2}%
$,\cite{Nohara99-QP-inVortex-Swave} while in the second, $T_{\beta\gamma}$ is
increased on doping such as in the case of NbW$_{x}$%
,\cite{Dasilva69-Nb-SpHeat} Li$_{2}$(Pd$_{1-x}$Pt$_{x}$)$_{3}$%
B,\cite{07-Li2T3B} and Li$_{x}$RhB$_{1.5}$.\cite{10-Li-Rh-B-JPSJ}

Finally, an anomaly in the electron DOS, such as an opening of a pseudogap,
would be coupled by a relatively strong electron-phonon interaction to the
phonon excitations in the same way as was described above for the case of
\textrm{Nb}$_{\mathrm{3}}$\textrm{Sn:}\cite{Stewart84-H-spHeat-A15} Similarly,
a phonon anomaly would be coupled to the electron excitations.

\section{Conclusions}

Variety of compounds exhibit a $\beta\gamma$-change; some are conventional
type-II superconductors, some are unconventional superconductors, while others
are\ normal intermetallics. In cases where both\ superconductivity and
$\beta\gamma$-anomalies are manifested, some compounds exhibit $T_{\beta
\gamma}<T_{c}$ while others $T_{\beta\gamma}>T_{c}$. The onset of $\beta
\gamma$-change can be sharp or smooth depending on material properties however
such an event is not related to a thermodynamic phase transition. The
strength, character, and trend of this $\beta\gamma$-change vary widely,
nonetheless, there is a systematic correlation between $\gamma$ and $\beta$.
Furthermore, this $\beta\gamma$-change can be influenced by perturbations such
as ASOC\ interaction and substitution but hardly by a variation in $N$($E_{F}%
$) or a magnetic field.

It was shown that this $\beta\gamma$-change is related to anomalies within the
phonon or electron dispersion relation. Assuming a Dirac-type anomaly in the
phonon and electron DOS curves, an analytical expression for the thermal
evolution of the total specific heat of the electron and phonon quasiparticles
was derived and was found to compare favorably with the studied experimental
curves. The term expressing the lattice contribution can be interpreted either
as a sum of a Debye and an Einstein mode or else as a Debye term with an
effective $T$-dependent $\theta_{D}$. The overall features of the resulting
$C/T$ $versus$ $T^{2}$ curve indicate (i) a manifestation of a break in the
slope at $T_{\beta\gamma}=0.2\left(  1\pm1/\sqrt{6}\right)  \theta_{L}$, (ii)
that the\ slope break is mostly determined by the phonon anomaly. The
correlation between $\theta_{D}$ and $\gamma$ is traced down to the influence
of the dielectric properties on the sound velocity of the low-temperature
acoustic phonons. Finally, the drop in the resistivity curve at $T_{\beta
\gamma}$ is shown to be caused by the same mechanism that gives rise to the
slope break in the $C/T$ $versus$ $T^{2}$ curve.

\begin{acknowledgments}
We acknowledge the partial financial support from the Brazilian agency CNPq.
\end{acknowledgments}

\bibliographystyle{apsrev}
\bibliography{ASOC-Unconv-Sup,borocarbides,intermetallic,MagClassic,massalami,notes,Sup-classic,To-Be-Published}

\end{document}